\documentclass{conm-p-l}

\newtheorem{theorem}{Theorem}[section]

\theoremstyle{definition}
\newtheorem{definition}[theorem]{definition}

\theoremstyle{remark}

\numberwithin{equation}{section}

\begin{document}

\title[Fortieth Anniversary of Extremal Projector Method]
{Fortieth Anniversary of Extremal Projector Method\\ for Lie
Symmetries}

\author{Valeriy N.\ Tolstoy}
\address{Institute of Nuclear Physics, Moscow State University,
Moscow 119992, Russia}
\email{tolstoy@nucl-th.sinp.msu.ru}
\thanks{This work was supported by Russian Foundation for Fundamental
Research, grant No.\ RFBR-02-01-00668 and INTAS-OPEN-03-51-3350.}

\subjclass[2000]{Primary 81R50; Secondary 17B37, 16W35}
\copyrightinfo{2005}{American Mathematical Society}

\keywords{Representation theory, projection operator, Clebsch-Gordan
coefficients, reduction algebras}

\begin{abstract}
A brief review of the extremal projector method for Lie symmetries
(Lie algebras and superalgebras as well as their quantum analogs)
is given. A history of its discovery and some simplest
applications are presented.
\end{abstract}

\maketitle
\section{Introduction}
In 1964 P.-O.~L\"owdin \cite{L}
first obtained an explicit expression of the extremal projection
operator for the Lie algebra $\mathfrak{sl}_2(\mathbb C)$. Later
such explicit formulas of the extremal projectors were found for
all finite-dimensional simple Lie algebras \cite{AS2,AST1,AST2,AST3},
classical Lie superalgebras \cite{T1}, infinite-dimensional affine
Kac-Moody algebras and superalgebras \cite{T4}, and also for their
quantum ($q$-de\-for\-med) analogs \cite{T5,KT1}. At present the
extremal projector method is a powerful and universal method for
solving many problems in representation theory. For
example, the method allows to classify irreducible modules, to
decompose them on submodules (e.g.~to analyze the structure of Verma
modules), to describe reduced (super)algebras (which are connected
with the reduction of a (super)algebra to a subalgebra), to construct
bases of representations (e.g.~the Gelfand-Tsetlin's type), to
develop the detailed theory of Clebsch-Gordan coefficients and
other elements of Wigner-Racah calculus (including compact
analytic formulas for these elements and their symmetry properties)
and so on.

In this paper we give a brief review of the extremal projector
method for Lie symmetries (Lie algebras and superalgebras as well
as their quantum analogs), namely, we provide a history of its
discovery and some simplest applications.

\section{Projection operators for finite and compact groups}

Let $G$ be a finite and compact group, and $T$ be its
representation in a linear space $V$, i.e. $g\mapsto T(g)$, $(g\in
G)$, where $T(g)$ are linear operators acting in $V$, satisfying
$T(g_1g_2)=T(g_1)T(g_2)$. The representation $T$ in $V$ is
irreducible if $\mathop{Lin}\{T(G)v\}=V$ for any nonzero vector
$v\in V$. An irreducible representation (IR) is denoted by an
additional upper index $\lambda$, $T^{\lambda}(g)$, and also
$V^{\lambda}$, or in matrix form:
$(T^{\lambda}(g))=(t^{\lambda}_{ij}(g))$ $(i,j=1,2,\ldots, n)$,
where $n$ is the dimension of $V^{\lambda}$.

It is well-known that the elements
\begin{equation}
P_{ij}^{\lambda}=\sum_{g\in G}T(g)t^{\lambda}_{ij}(g)\label{fg1}
\end{equation}
are projection operators for the finite group $G$, i.e.~they
satisfy the following properties:
\begin{eqnarray}
P_{ij}^{\lambda}P_{kl}^{\lambda'}&=&\delta_{\lambda\lambda'}
\delta_{jk} P_{il}^{\lambda},
\label{fg2}
\\[7pt]
(P_{ij}^{\lambda})^*&=&P_{ji}^{\lambda},
\label{fg3}
\end{eqnarray}
where $^*$ is Hermitian conjugation.

The projection operators for a compact group $G$ are modified as
follows:
\begin{equation}
P_{ij}^{\lambda}=\int\limits_{g\in G}T(g)t^{\lambda}_{ij}(g)dg~.
\label{fg4}
\end{equation}
In the case $G=SO(3)$ (or $SU(2)$) we have
\begin{equation}
P_{mm'}^{j}=\int_{}\,
T(\alpha,\beta,\gamma)\,D^j_{mm'}(\alpha,\beta,\gamma)\sin\beta
\,d\alpha\, d\beta\, d\gamma~, \label{fg5}
\end{equation}
where $\alpha,\beta,\gamma$ are the Euler angles and
$D^j_{mm'}(\alpha,\beta,\gamma)$ is the Wigner $D$-function. The
projection operator $P^j:=P^j_{jj}$ is called the projector on the
highest weight $j$.

Thus we see that the projection operators in the form (\ref{fg1})
or (\ref{fg4}) require explicit expressions for the operator
function $T(g)$, the matrix elements of IRs $t^{\lambda}_{ij}(g)$,
and also (in the case of a compact group) the $g$-invariant
measure $dg$. In the case of an arbitrary compact group $G$ these
expressions lead to several problems.
\smallskip

\section{L\"owdin-Shapiro extremal projector for
the angular momentum\\ Lie algebra}

The angular momentum Lie algebra $\mathfrak{so}(3)$ ($\simeq
\mathfrak{su}(2)$) is generated by the three elements (generators)
$J_{+}$, $J_{-}$ and $J_{0}$ with the defining relations:
\begin{equation}
\begin{array}{rcl}
[J_0,J_{\pm}] &\!\! =\!\!& \pm J_{\pm},\qquad
[J_{+},J_{-}]=2J_0,
\\[7pt]
J_{\pm}^*& \!\! =\!\!& J_{\mp}~,\qquad\qquad\;\; J_0^*\,=\,J_0~.
\label{ls1}
\end{array}
\end{equation}
The Casimir element ${\bf C}_2$ of the angular momentum Lie algebra
(or square of the angular momentum  ${\bf J}^2$) is given by:
\begin{equation}
{\bf C}_2\,\equiv\,{\bf J}^2 \,=\, \frac{1}{2}\Bigl(J_+^{}J_-^{}
+J_-^{}J_+^{}\Bigr)+J_0^2\,=\, J_-^{}J_+^{}+J_0^{}(J_0^{}+1)~,
\label{ls2}
\end{equation}
\begin{equation}
[J_i~,{\bf J}^2]\,=\,0~.
\label{ls3}
\end{equation}
Let $\{|jm\bigr>\}$ be the canonical basis of
$\mathfrak{su}(2)$-IR corresponding to the spin $j$ (wave
functions with definite $j$ and its projection $m$ ($m=-j,
-j+1,\dots,j)$, for example, spherical harmonics, $|jm\bigr>\equiv
Y_m^j$). These basis functions satisfy the relations:
\begin{equation}
\begin{array}{rcl}
{\bf J}^2|jm\bigr>&\!\!=\!\!&j(j+1)|jm\bigr>,\quad
J_0|jm\bigr>\;=\;m|jm\bigr>~,
\\[7pt]
J_{\pm}|jm\bigr>&\!\!=\!\!&\sqrt{(j\mp m)(j\pm m+1)}
\;|jm\pm1\bigr>~. \label{ls4}
\end{array}
\end{equation}
The vectors $|jm\bigr>$ can be represented as follows:
\begin{equation}
|jm\bigr>=F_{\!m;j}^{\,j}|jj\bigr>~,
\label{ls5}
\end{equation}
where
\begin{equation}
F_{\!m;j}^{\,j}\!=
\sqrt{\frac{(j+m)!}{(2j)!(j-m)!}}\,J_{-}^{j-m}\quad\
\Bigl(\!F_{\!j;m}^{\,j}\!:=(F_{\!m;j}^{\,j})^*\!=
\sqrt{\frac{(j+m)!}{(2j)!(j-m)!}}\,J_{+}^{j-m}\Bigr), \label{ls6}
\end{equation}
and $|jj\bigr>$ is the highest weight vector, i.e.
\begin{equation}
J_{+}^{}|jj\bigr>=0~.
\label{ls7}
\end{equation}
It is obvious that the projector $P^j$ on the highest weight,
(\ref{fg5}), satisfies the relations:
\begin{equation}
J_{+}^{}P^j\,=\,P^jJ_{-}^{}\,=\,0,\qquad (P^j)^2\,=\,P^j~.
\label{ls8}
\end{equation}

An associative polynomial algebra of the generators $J_{\pm}$,
$J_0$ is called the universal enveloping algebra of the angular
momentum Lie algebra and it is denoted by $U(\mathfrak{so}(3))$ (or
$U(\mathfrak{su}(2))$. The following proposition holds.

``No-go theorem": {\it No nontrivial solution of the equations}
\begin{equation}
J_{+}^{}P\,=\,PJ_{-}^{}\,=\,0
\label{ls9}
\end{equation}
{\it exists in $U(\mathfrak{su}(2))$, i.e.~a unique solution of
these equations for $P\in U(\mathfrak{su}(2))$ is trivial
$P\equiv0$}.\,\footnote{~A general mathematical statement of this
theorem reads as follows: ``{\it The universal enveloping algebra
$U(\mathfrak{g})$ of a simple Lie algebra $\mathfrak{g}$ has no
zero divisors.}"}

Thus the theorem states that the projector $P^j$ does not exist in
the form of a polynomial of the generators $J_{\pm}$, $J_0$. This
no-go theorem was well known to mathematicians, but we can assume
that it was not known to most physicists.

In 1964, exactly 40 years ago, the Swedish physicist and chemist
P.-O.~L\"owdin\,\footnote{~Per-Olov L\"owdin was
born in 1916 in Uppsala, Sweden, and died in 2000 (see
http://\linebreak[0]www.quantum-chemistry-history.com/Lowdin1.htm).},
who probably did not know the no-go theorem, published a paper in
the journal {\it Rev.\,Mod.\,Phys.} \cite{L}, in which he considered the
following operator:
\begin{equation}
P^j\,:=\,\prod_{j'\neq j}\frac{{\bf J}^2-j'(j'+1)}
{j(j+1)-j'(j'+1)}~.
\label{ls10}
\end{equation}
This element has the following properties. Let $\Psi_{m=j}$ be an
arbitrary eigenvector of the operator $J_0$:
\begin{equation}
J_0\Psi_{m}\,=\,m\Psi_{m}~.
\label{ls11}
\end{equation}
Due to completeness of the basis formed by the vectors
$\{|jm\bigr>\}$ (for all possible spins $j$ and their projections
$m$) the following expansion holds:\,\footnote{~Here we consider a
multiplicity free case.}
\begin{equation}
\Psi_{m}\,=\,\sum_{j'}C_{j'}|j'm\bigr>~,
\label{ls12}
\end{equation}
and it is obvious that
\begin{equation}
P^j\Psi_{m=j}\,=C_j|jj\bigr>~.
\label{ls13}
\end{equation}
From here we obtain the following properties of the element
(\ref{ls10}):
\begin{equation}
J_{+}^{}P^j\,=\,P^jJ_{-}^{}\,=\,0~, \label{ls14}
\end{equation}
\begin{equation}
[J_{0}^{},\,P^j]\,=\,0~,\qquad (P^j)^2=P^j~, \label{ls15}
\end{equation}
provided that the left and right sides of these equalities act on
vectors with a definite projection of angular momentum $m=j$.
Therefore the element (\ref{ls10}) is the projector on the highest
weight.

After rather complicated calculations L\"owdin reduced
the operator (\ref{ls10}) to the following form:
\begin{equation}
P^j\,=\,\sum_{n\ge0}\,\frac{(-1)^n(2j+1)!}{n!(2j+n+1)!}\,
J_-^nJ_+^n~.
\label{ls16}
\end{equation}

One year later, in 1965, another physicist, J.~Shapiro from USA,
published a paper in {\it J.\,Math.\,Phys.}\ \cite{Sh}, in which he
stated:
``Let us forget the initial expression (\ref{ls10}) and consider
the defining relations (\ref{ls14}) and (\ref{ls15}), where $P^j$
has the following ansatz:
\begin{equation}
P^j\,=\,\sum_{n\ge0}\,C_n(j)\,J_-^nJ_+^n~.\;\; "
\label{ls17}
\end{equation}
Substituting this expression in (\ref{ls14}) we directly obtain
the formula (\ref{ls16}).

We can remove the upper index $j$ in $P^j$ if we replace
$j\rightarrow J_0$:
\begin{eqnarray}
P&\!\!=&\!\!\sum_{n\ge0}\,\frac{(-1)^n}{n!}\;
\varphi_{n}(J_0)\,J_-^nJ_+^n~, \label{ls18}
\\[7pt]
\varphi_{n}(J_0)&\!\!=&\!\!\prod_{k=1}^{n}(2J_0+k+1)^{-1}~.
\label{ls19}
\end{eqnarray}
The element $P$ is called the extremal projector. If $\Psi$ is an
arbitrary function
\begin{equation}
\Psi\,=\,\sum_{j,m}C_{j,m}|jm\bigr>~,
\label{ls20}
\end{equation}
then
\begin{equation}
P\Psi\,=\,\sum_{j}C_{j,j}|jj\bigr>~.
\label{ls21}
\end{equation}
The extremal projector $P$ does not belong to
$U(\mathfrak{su}(2))$ but it belongs to some extension of the
universal enveloping algebra. Let us determine this extension.

Consider the formal Taylor series
\begin{equation}
\sum_{n,k\ge0}C_{n,k}(J_0)\,J_-^nJ_+^k\ ,
\label{ls22}
\end{equation}
where $C_{n,k}(J_0)$ are rational functions of the Cartan element $J_0$ and
provided that for each series there exists a natural number $N$ for which
\begin{equation}
|n-k|\le N\ .
\label{ls23}
\end{equation}

Let $TU(\mathfrak{su}(2))$ be the linear space of such formal
series. We can show  that {\it $TU(\mathfrak{su}(2))$ is an
associative algebra with respect to the multiplication of
formal series}. The associative algebra $TU(\mathfrak{su}(2))$ is
called the {\it Taylor extension} of $U(\mathfrak{su}(2))$. It is
obvious that $TU(\mathfrak{su}(2))$ contains
$U(\mathfrak{su}(2))$.

{\it Remark.} The restriction (\ref{ls23}) is important. Consider
two series:
\begin{equation}
x_1:=\sum_{k\ge0}\,J_+^k~,\qquad x_2:=\sum_{n\ge0}\,J_-^n~.
\label{ls24}
\end{equation}
Their product is reduced to the form (\ref{ls22})
\begin{equation}
x_1x_2=\sum_{n,k\ge0}\Delta_{n,k}(J_0)\,J_-^nJ_+^k~, \label{ls25}
\end{equation}
where $\Delta_{n,k}(J_0)$ is not any rational function of $J_0$,
and moreover it is a generalized function of $J_0$.

The extremal projector (\ref{ls18}) belongs to the Taylor
extension $TU(\mathfrak{su}(2))$. The\-refore L\"owdin
and Shapiro found a solution of the equations (\ref{ls14}), not in
the space $U(\mathfrak{su}(2))$, but in its extension
$TU(\mathfrak{su}(2))$.

Later Shapiro tried to generalized the obtained formula
(\ref{ls18}) to the case of $\mathfrak{su}(3)$
($\mathfrak{u}(3)$). The Lie algebra $\mathfrak{u}(3)$ is
generated by 9 elements $e_{ik}$ $(i,k=1,2,3)$ with the relations:
\begin{equation}
[e_{ij},\,e_{kl}]\,=\,\delta_{jk}e_{il}-\delta_{il}e_{kj}~,
\qquad e_{ij}^*\,=\,e_{ij}~.
\label{ls26}
\end{equation}
Shapiro considered the following ansatz for $P$:
\begin{equation}
P:=\sum_{n_i,m_i\ge0}C_{n_{i},m_{i}}(e_{11},e_{22},e_{33})
\,e_{21}^{n_1}e_{31}^{n_2}e_{32}^{n_3}e_{12}^{m_1}
e_{13}^{m_2}e_{23}^{m_3}
\label{ls27}
\end{equation}
and he used the equations
\begin{eqnarray}
e_{ij}^{}P&\!\!=&\!\!Pe_{ji}^{}\,=\,0\quad (i<j)~,
\label{ls28}
\\[7pt]
[e_{ii},\,P]&\!\!=&\!\!0\quad (i=1,2,3)~.
\label{ls29}
\end{eqnarray}
From the last equation it follows that
\begin{equation}
\left\{\begin{array}{rcl}
n_1+n_2&\!\!=&\!\!m_1+m_2~,
\\[3pt]
n_2+n_3&\!\!=&\!\!m_2+m_3~.
\end{array}
\right.
\label{ls30}
\end{equation}
Under the conditions (\ref{ls30}) the expression (\ref{ls27})
belongs to $TU(\mathfrak{su}(3))$.  A system of equations for the
coefficients $C_{n_{i},m_{i}}(e_{11},e_{22},e_{33})$ was found to
be too complicated, and Shapiro failed to solve this system.

In 1968 R.M.~Asherova and Yu.F.~Smirnov \cite{AS1} made the first
important step in order to obtain an explicit formula for the
extremal projector for $\mathfrak{u}(3)$. They proposed to act
with $P$ described by the Shapiro ansatz (\ref{ls27}) on the
extremal projector of the subalgebra $\mathfrak{su}(2)$ generated
by the elements $e_{23},\;e_{32},\;e_{22}-e_{33}$,
\begin{eqnarray}
P_{23}&\!\!=&\!\!\sum_{n\ge0}\,\frac{(-1)^n}{n!}\;
\varphi_{n}(e_{22}-e_{33})\,e_{32}^ne_{23}^n~,\label{ls31}
\end{eqnarray}
where
\begin{eqnarray}
\varphi_{n}(e_{22}-e_{33})&\!\!=&\!\!
\prod_{k=1}^{n}(e_{22}-e_{33}+k+1)^{-1}~. \label{ls32}
\end{eqnarray}
Since $e_{23}P_{23}=0$, therefore we obtain the following form
for $P(\mathfrak{su}(3))$:
\begin{equation}
P=\sum_{n_{i}\ge0}C_{n_{1},n_{2},n_{3}}(e_{11},e_{22},e_{33})
\,e_{21}^{n_1}e_{31}^{n_2-n_3}e_{32}^{n_3}e_{12}^{n_1-n_3}
e_{13}^{n_2}\,P_{23}~.
\label{ls33}
\end{equation}
In this case the system of equations for the coefficients
$C_{n_i}(e_{ii})$ is simpler, and it was solved. However, the
explicit expressions for the coefficients $C_{n_i}(e_{ii})$ are
rather complicated. The next simple idea \cite{T} was to act on
the expression (\ref{ls33}) (from the left side) by the extremal
projector of the subalgebra $\mathfrak{su}(2)$ generated by the
elements $e_{12},\,e_{21},e_{11}-e_{22}$. As a result we obtain
the following simple form for $P:=P(\mathfrak{su}(3))$:
\begin{equation}
P=P_{12}\Bigl(\sum_{n\ge0}C_{n}(e_{11}-e_{33})
\,e_{31}^{n}e_{13}^{n}\Bigr)P_{23}~.
\label{ls34}
\end{equation}
The final formula is
\begin{equation}
P=P_{12}P_{13}P_{23}~,
\label{ls35}
\end{equation}
where
\begin{eqnarray}
P_{ij}&\!\!=&\!\!\sum_{n\ge0}\,\frac{(-1)^n}{n!}\;
\varphi_{n}(e_{ii}-e_{jj})\,e_{ji}^ne_{ij}^n\quad (i<j)~,
\label{ls36}
\\[7pt]
\varphi_{n}(e_{ii}-e_{jj})&\!\!=&\!\!
\prod_{k=1}^{n}(e_{ii}-e_{jj}+k+j-i)^{-1}~. \label{ls37}
\end{eqnarray}
It was found that this formula is fundamental. In the next section
we give the explicit formula of the extremal projector for all
finite-dimensional simple Lie algebras.

\section{Extremal projector for simple Lie algebras}

Let $\mathfrak{g}$ be a finite-dimensional simple Lie algebra and
$\Delta_+(\mathfrak{g})$ be its positive root system. A
generalization of the formulas (\ref{ls35})--(\ref{ls37}) to the
case of $\mathfrak{g}$ is connected with the notion of normal
ordering in the system $\Delta_+(\mathfrak{g})$.
\begin{definition}
We say that the system $\Delta_+(\mathfrak{g})$ is in normal
ordering if each composite (not simple) root
$\gamma\!=\!\alpha\!+\!\beta$
($\alpha,\beta,\gamma\in\Delta_+(\mathfrak{g})$) is written
between its constituents $\alpha$ and $\beta$.
\end{definition}
{\it Remarks}. {\it (i)} We can show that the normal ordering in
the system $\Delta_+(\mathfrak{g})$ exists (see \cite{T6}). {\it
(ii)} For classical simple Lie superalgebras the normal ordering
is defined for a reduced root system
$\underline{\Delta}_+(\mathfrak{g})$ \cite{T1}, and for
infinite-dimensional affine Kac--Moody algebras and superalgebras
the definition of the normal ordering is modified \cite{T4,T5,T7}.

The normally ordered system $\Delta_+(\mathfrak{g})$ is denoted by
the symbol $\vec{\Delta}_+(\mathfrak{g})$. Let $e_{\pm\gamma}^{}$,
$h_{\gamma}^{}$ be Cartan-Weyl root vectors normalized by the
condition
\begin{equation}
[e_{\gamma}^{},e_{-\gamma}^{}]=h_{\gamma}^{}~.
\label{ep1}
\end{equation}
\begin{theorem}
The equations
\begin{equation}
e_{\gamma}^{}P=Pe_{-\gamma}^{}=0\quad(\forall\;
\gamma\in\Delta_+(\mathfrak{g}))~,
\qquad P^{2}=P \label{ep2}
\end{equation}
have a unique nonzero solution in the space of the Taylor
extension $T_{q}(\mathfrak{g})$ and this solution has the form
\begin{equation}
P=\prod_{\gamma\in\vec{\Delta}_+(\mathfrak{g})}\!\!
P_{\gamma}^{}~, \label{ep3}
\end{equation}
where the elements $P_{\gamma}$ are defined by the formulae
\begin{equation}
P_{\gamma}^{}=\sum_{n\geq0}\frac{(-1)^{n}}
{n!}\,\varphi_{\gamma,n}^{}e_{-\gamma}^{\,m}\,e_{\gamma}^{m}~,
\label{ep4}
\end{equation}
\begin{equation}
\varphi_{\gamma,n}^{}=\prod\limits_{k=1}^{n}\Bigl(h_{\gamma}^{}
+(\rho,\gamma)+\frac{1}{2}(\gamma,\gamma)k\Bigr)^{-1}~.
\label{ep5}
\end{equation}
Here $\rho$ is the half-sum of all positive roots.
\end{theorem}
In the next Sections 5--8 we consider some simplest applications
of the extremal projectors for the Lie algebras $\mathfrak{su}(2)$
and $\mathfrak{su}(3)$.

\section{Clebsch-Gordan coefficients for the angular momentum Lie
algebra}

Let $\{|j_im_i\bigr>\}$ be wave functions of two systems
(canonical bases of two IRs of the angular momentum Lie algebra)
with spins $j_i$ $(i=1,2)$. Then
$\{|j_1m_1\bigr>|j_2m_2\bigr>\}$ is the uncoupled (tensor) basis
in the representation $j_1\otimes j_2$ for
$\mathfrak{su}(2)\otimes\mathfrak{su}(2)$. In this representation
there is another basis $|j_1j_2\!:\!j_3m_3\bigr>$ which is called
coupled with respect to $\mathfrak{su}(2)$ (which is a diagonal
embedding in $\mathfrak{su}(2)\otimes\mathfrak{su}(2)$). We can
extend the coupled basis in terms of the uncoupled one:
\begin{equation}
\bigl|j_1j_2\!:\!j_3m_3\bigr>=\sum_{m_1,m_2}^{}
\bigl(j_1m_1\,j_2m_2|j_3m_3\bigr)\bigr|j_1m_1\bigr>
\bigr|j_2m_2\bigr>~, \label{cg1}
\end{equation}
where matrix elements $\bigl(j_1m_1\,j_2m_2|j_3m_3\bigr)$ are called
Clebsch-Gordan coefficients (CGC).

One can show that CGC can be presented in the form:
\begin{equation}
\bigl(j_1m_1\,j_2m_2\bigr|j_3m_3\bigr)=\frac{\langle j_1m_1|
\langle j_2m_2|P_{\!m_3;j_3}^{j_3}|j_1j_1\rangle|
j_2j_3-j_1\rangle} {\sqrt{\langle j_1j_1|\langle j_2\,j_3-j_1|
P_{\!j_3;j_3}^{j_3}|j_1j_1\rangle|j_2\,j_3-j_1\rangle}}
\label{cg2}
\end{equation}
where  $P_{m_3;m_3'}^{j_3}$ is a general projection operator
which is connected with the extremal projector as follows:
\begin{equation}
P_{\!m_3;m_3'}^{\,j_3}:=F_{\!m_3;j_3}^{\,j_3}\,P^{j_3}\,
F_{\!j_3;m_3'}^{\,j_3}~, \label{cg3}
\end{equation}
and it is constructed from the generators of the coupled system,
$J_i(3)=J_i(1)+J_i(2)\equiv\Delta(J_{i})=J_i\otimes1+1\otimes
J_i$,  ($i=\pm,0$).

Using the explicit formulas (\ref{ls6}) and (\ref{ls16}) one can
easily obtain the final formula of $\mathfrak{su}(2)$-CGC:
\begin{equation}
\begin{array}{l}
\bigl(j_1m_1\,j_2m_2\bigr|j_3m_3\bigr)=\delta_{m_1\!+m_2,m_3}^{}
\\[7pt]
\qquad\mbox{\Large$\times{\sqrt{\frac{(2j_3+1)(j_2-m_2)!(j_3+m_3)!
(j_1+j_2+j_3+1)!(j_1+j_2-j_3)!(j_1-j_2+j_3)!}
{(j_1+m_1)!(j_1-m_1)!(j_2+m_2)!(j_3-m_3)!(-j_1+j_2+j_3)!}}}$}
\\[12pt]
\qquad\mbox{\Large$\times\sum\limits_{n}\frac{(-1)^{j_1+j_2-j_3-n}(2j_2-n)!
(j_1+j_2-m_3-n)!}{n!(j_2-m_2-n)!(j_1+j_2-j_3-n)!(j_1+j_2+j_3+1-n)!}$}~.
\end{array}
\label{cg4}
\end{equation}
This formula was obtained by Shapiro \cite{Sh}, and it allows to
obtain all classical symmetry properties of CGC: permutations and
conjugations.

\setcounter{equation}{0}
\section{Gelfand-Tsetlin basis for $\mathfrak{su}(3)$}

The Lie algebra $\mathfrak{su}(3)$ is generated by the elements
$e_{ij}$ $(i,j=1,2,3)$ provided $e_{11}+e_{22}+e_{33}=0$ with the
relations (\ref{ls26}). Let $(\lambda\mu)$ be a finite-dimensional
irreducible representation (IR) of $\mathfrak{su}(3)$ with the
highest weight $(\lambda\mu)$ ($\lambda$ and $\mu$ are nonnegative
integers). The highest weight vector, denoted by the symbol
$\bigr|(\lambda\mu)h\bigl>$, satisfies the relations
\begin{equation}
\begin{array}{rcl}
(e_{11}-e_{22})\bigr|(\lambda\mu)h\bigl> &\!\!=\!\!&
\lambda\bigr|(\lambda\mu)h\bigl>~,
\\[7pt]
(e_{22}-e_{33})\bigr|(\lambda\mu)h\bigl>&\!\!=\!\!&
\mu\bigr|(\lambda\mu)h\bigl>~,
\\[7pt]
e_{ij}^{}\bigr|(\lambda\mu)h\bigl> &\!\!=\!\!&
0\quad (i<j)~.
\end{array}
\label{gt1}
\end{equation}
The labelling of other basis vectors in IR $(\lambda\mu)$ depends
upon the choice of  subalgebras of $\mathfrak{su}(3)$ (or in another
words, depends upon which reduction chain from $\mathfrak{su}(3)$
to the subalgebras is chosen). Here we use the Gelfand-Tsetlin
reduction chain:
\begin{equation}
\mathfrak{su}(3)\supset u_{Y}^{}(1)\otimes
\mathfrak{su}_{T}^{}(2)\supset u_{T_{0}}^{}(1)~, \label{gt2}
\end{equation}
where the subalgebra $\mathfrak{su}_{T}^{}(2)$ is generated by the
elements
\begin{equation}
T_{+}:=e_{23}^{}~,\qquad T_{-}:=e_{32}^{}~,\qquad
T_{0}:=\mbox{\large$\frac{1}{2}$}(e_{22}^{}-e_{33}^{})~,
\label{gt3}
\end{equation}
the subalgebra $u_{T_0}^{}(1)$ is generated by ${T_{0}}$, and
$u_{Y}^{}(1)$ is generated by ${Y}$\,\footnote{~In elementary
particle theory the subalgebra $\mathfrak{su}_{T}^{}(2)$ is called
the T-spin algebra and the element ${Y}$ is called the hypercharge
operator.}
\begin{equation}
Y=-\mbox{\large$\frac{1}{3}$}\bigl(2e_{11}^{}-e_{2}^{}-e_{3}^{}\bigr)~.
\label{gt4}
\end{equation}
In the case of the reduction chain (\ref{gt2}) the basis vectors
of IR $(\lambda\mu)$ are denoted by
\begin{equation}
\bigl|(\lambda\mu)jtt_{z}\bigr>~. \label{gt5}
\end{equation}
Here the set $jtt_{z}$ characterizes the hypercharge $Y$ and the
T-spin and its projection:
\begin{equation}
\begin{array}{rcl}
{Y}\bigl|(\lambda\mu)jtt_{z}\bigr>&\!\!=\!\!&{y}
\bigl|(\lambda\mu)jtt_{z}\bigr>~,
\quad{T_0}\bigl|(\lambda\mu)jtt_{z}\bigr>\;=\;{t_z}
\bigl|(\lambda\mu)jtt_{z}\bigr>~,
\\[12pt]
T_{\pm}\bigl|(\lambda\mu)jtt_{z}\bigr>&\!\!=\!\!&\sqrt{(t\mp t_z)
(t\pm t_z\!+\!1)}\bigl|(\lambda\mu)jtt_{z}\!\pm\!1\bigr>~,
\end{array}
\label{gt6}
\end{equation}
where the parameter $j$ is connected with the eigenvalue $y$ of
the operator $Y$ as follows:
$y=-\mbox{\large$\frac{1}{3}$}\bigl(2\lambda+\mu\bigr)+2j$. It is
not hard to show (see \cite{PST1,AST5}) that the orthonormalized
vectors (\ref{gt6}) can be represented in the following form:
\begin{equation}
\bigr|(\lambda\mu)jtt_{z}\bigr>= {F}_{\!-}^{(\lambda\mu)}(jtt_z)
\bigr|(\lambda\mu)h\bigr>:=N^{(\lambda\mu)}_{\,jt}
P^{\,t}_{\!t_{z};t}\;e_{31}^{j+\frac{1}{2}\mu-t}
e_{21}^{j-\frac{1}{2}\mu+t}\bigr|(\lambda\mu)h\bigr>~,\label{gt7}
\end{equation}
where $P^{\,t}_{\!t_{z};t_z'}$ is the general projection operator
of the type (\ref{cg3}) for the Lie algebra
$\mathfrak{su}_{T}^{}(2)$, and the normalization factor
$N^{(\lambda\mu)}_{jt}$ has the form
\begin{equation}
N^{(\lambda\mu)}_{\,jt}=\mbox{\Large$\left(\frac{(\lambda+
\frac{1}{2}\mu-j+t+1)!(\lambda+\frac{1}{2}\mu-j-t)!
(\frac{1}{2}\mu+j+t+1)!(\frac{1}{2}\mu-j+t)!}{\lambda!\mu!
(\lambda+\mu+1)!(j+\frac{1}{2}\mu-t)!
(j-\frac{1}{2}\mu+t)!(2t+1)!}\right)^{\frac{1}{2}}\!$}.
\label{gt8}
\end{equation}
The quantum numbers $jt$ take all nonnegative integers and
half-integers such that the sum
$\mbox{\large$\frac{1}{2}$}\mu+j+t$ is an integer and they are
subjected to the constraint
\begin{equation}
\left\{\begin{array}{rcl}
\frac{1}{2}\mu+j-t&\!\!\ge\!\!&0~,\quad\;-\frac{1}{2}\mu+j+t\ge0~,
\\[3pt]
\frac{1}{2}\mu-j+t&\!\!\ge\!\!&0~,
\qquad\frac{1}{2}\mu+j+t\le\lambda+\mu~.
\end{array}
\right. \label{gt9}
\end{equation}
For every fixed $t$ the projection $t_z$ takes the values
$t_z=-t,-t+1,\ldots,t-1,t$.

The explicit form (\ref{gt7}) of the basis vectors
$\{\bigl|(\lambda\mu)jtt_{z}\bigr>\}$ allows to calculate easily
the actions of the generators $e_{ij}^{}$ (see \cite{PST1,AST5}).

\setcounter{equation}{0}
\section{Tensor form of the $\mathfrak{su}(3)$
projection operator}

It is obvious that the extremal projector of $\mathfrak{su}(3)$
can be presented in the form
\begin{equation}
P(\mathfrak{su}(3))=P(\mathfrak{su}^{}_T(2))
\bigl(P_{12}^{}P_{13}^{}\bigr)P(\mathfrak{su}^{}_T(2)).
\label{tf1}
\end{equation}
Now we present the middle part of (\ref{tf1}) in terms of the
$\mathfrak{su}_T(2)$ tensor operators. To this end, we substitute
the explicit expression for the factors $P_{12}$ and $P_{13}$, and
combine monomials  $e_{21}^{n}e_{31}^{m}$ and
$e_{12}^{n}e_{13}^{m}$. After some manipulations with sums we
obtain the following expression for the extremal projection
operator $P\!:=\!P(\mathfrak{su}(3))$ in terms of tensor operators:
\begin{equation}
P=P(\mathfrak{su}^{}_T(2)) \Bigr(\sum\limits_{jj_z} A_{jj_z}
\tilde{R}^{j}_{j_z}R^{j}_{\!j_z}\Bigl)P(\mathfrak{su}^{}_T(2))
\label{tf2}
\end{equation}
where
\begin{equation}
A_{jj_z}=\frac{(-1)^{3j}\varphi_{12}^{}
(\varphi_{12}^{}+j+j_z-1)!(\varphi_{13}^{})!}
{(2j)!(\varphi_{12}^{}+2j)!(\varphi_{13}^{}+j+j_z)!}~, \label{tf3}
\end{equation}
$\varphi_{1i+1}:=e_{11}\!-e_{i+1i+1}\!+i$, ($i=1,2$), and
\begin{eqnarray}
\tilde{R}^{j}_{j_z}&\!\!=&\!\!
\sqrt{\frac{(2j)!}{(j-j_{z})!(j+j_{z})!}}
\,e_{21}^{j+j_{z}}\,e_{31}^{j-j_{z}}, \label{tf4}
\\[7pt]
R^{j}_{\!j_z}&\!\!=&\!\!\sqrt{\frac{(2j)!}{(j-j_{z})!(j+j_{z})!}}
\,e_{12}^{j-j_{z}}\,e_{\!\!13}^{j+j_{z}}. \label{tf5}
\end{eqnarray}
The elements $R_{j_z}^{j}$ and $\tilde{R}_{j_z}^{j}$ are
irreducible tensor components, i.e.~they satisfy the relations
\begin{equation}
[T_{i}^{},R_{j_z}^{j}]=\sum\limits_{j'_z}
\bigl<jj'_z\bigl|T_{i}^{}\bigr|jj_z\bigr>R_{j'_z}^{j}~.
\label{tf6}
\end{equation}

Below we assume that the $\mathfrak{su}(3)$ extremal projection
operator (\ref{tf2}) acts in a weight space with the weight
$(\lambda\mu)$ and in this case the symbol $P$ is supplied with
the index $(\lambda\mu)$, $P^{(\lambda\mu)}$, and all the Cartan
elements $e_{ii}-e_{i+1i+1}$ on the right side of (\ref{tf2}) are
replaced by the corresponding weight components $\lambda$ and
$\mu$.

Now we multiply the projector $P^{(\lambda\mu)}$ from the left
side by the lowering operator ${F}_{\!-}^{(\lambda\mu)}(jtt_z)$
and from the right side by the rising operator
$\bigl({F}_{\!-}^{(\lambda\mu)}(jtt_z)\bigr)^*$, and we finally
find the tensor form of the general $\mathfrak{su}(3)$ projection
operator:
\begin{equation}
P^{(\lambda\mu)}_{\!\!jtt_z;j't't'_z}\!=
\sum\limits_{j''t''}B_{j''t''}^{(\lambda\mu)}\;{\widetilde{\bf
I\!R}}^{j+j''}_{tt_z,t''t''}\; {\bf
I\!R}^{j''\!+j'}_{\!t''t'',t't'_z}~, \label{tf7}
\end{equation}
were the coefficients $B_{j'\!'t'\!'}^{(\lambda\mu)}$ are given
by
\begin{equation}
\begin{array}{l}
B_{j''t''}^{(\lambda\mu)}\;=\;
\mbox{\LARGE$\frac{(-1)^{2j+j'\!+j''\!-t'+t''}
(\lambda+1)(\mu+1)(\lambda+\mu+2)}
{(\lambda+\frac{1}{2}\mu+j''\!+t''\!+2)!
(\lambda+\frac{1}{2}\mu+j''\!-t''+1)!(2j'')!}$}\;\;
{\displaystyle\left\{{j\atop t''}\; {j''\atop
t}\;{j\!+\!j''\atop\frac{1}{2}\mu}\right\}}
\\[14pt]
\qquad\quad\times{\displaystyle\left\{{j'\atop t''}\;
{j''\atop t'}\;{j'\!+\!j''\atop\frac{1}{2}\mu}\right\}}
\sqrt{(\lambda+\frac{1}{2}\mu-j+t+1)!
(\lambda+\frac{1}{2}\mu-j-t)!}
\\[12pt]
\qquad\quad\times\mbox{\LARGE$\left(\frac{
(\lambda+\frac{1}{2}\mu-j'\!+t'\!+1)!
(\lambda+\frac{1}{2}\mu-j'\!-t')!(2j+2j''\!+1)(2j'+2j''\!+1)}
{(2j)!(2j')!(2t+1)(2t'+1)} \right)^{\frac{1}{2}}$}.
\end{array}
\label{tf8}
\end{equation}
The operators ${\widetilde{\bf I\!R}}^{j+j''}_{tt_z,t''t''}$ and
${\bf I\!R}^{j''\!+j'}_{\!t''t'',t't'_z}$ are given by
\begin{equation}
{\bf I\!R}^{j}_{tt_z;t't'_z}:=\sqrt{(2t+1)}\;
\sum\limits_{j_zt'\!'_{\!z}} \bigl(j
j_z\,tt''_z\bigl|t't'_z\bigr)\,R_{j_z}^{j}
P^{t}_{\!t'\!'_{\!z},t_z}~. \label{tf9}
\end{equation}
The formula (\ref{tf7}) is the key for the calculation of
$\mathfrak{su}(3)$-Clebsch-Gordan coefficients.

\setcounter{equation}{0}
\section{General form of Clebsch-Gordan coefficients for
$\mathfrak{su}(3)$}

For convenience we introduce the short notations
$\Lambda:=(\lambda\mu)$ and $\gamma:=jtt_{z}$, and therefore the
basis vector $\bigl|(\lambda\mu)jtt_z\bigr>$ will be denoted by
$\bigl|\Lambda\gamma\bigr>$. Let $\{|\Lambda_i\gamma_i\bigr>\}$
denote bases of two IRs $\Lambda_i$ $(i=1,2)$. Then
$\{|\Lambda_1\gamma_1\bigr>|\Lambda_2\gamma_2\bigr>\}$ form a
basis in the representation $\Lambda_1\otimes \Lambda_2$ of
$\mathfrak{su}(3)\otimes \mathfrak{su}(3)$. In such a
representation there is another coupled basis
$|\Lambda_1\Lambda_2\!:s\Lambda_3\gamma_3\bigr>$, where the index
$s$ classifies the multiplicity of the representations $\Lambda_3$. We
can expand the coupled basis in terms of the tensor (``uncoupled")
basis $\{|\Lambda_1\gamma_1\bigr>|\Lambda_2\gamma_2\bigr>\}$:
\begin{equation}
\bigl|\Lambda_1\Lambda_2\!:s\Lambda_3\gamma_3\bigr>=
\sum_{\gamma_1,\gamma_2}^{}\bigl(\Lambda_1\gamma_1\,
\Lambda_2\gamma_2\bigl|s\Lambda_3\gamma_3\bigr)\,
\bigr|\Lambda_1\gamma_1\bigr>\bigr|\Lambda_2\gamma_2\bigr>~,
\label{CGC1}
\end{equation}
where the matrix element
$\bigl(\Lambda_1\gamma_1\,\Lambda_2\gamma_2|s\Lambda_3\gamma_3\bigr)$
is the Clebsch-Gordan coefficient of $\mathfrak{su}(3)$.

We can show that any CGC of $\mathfrak{su}(3))$ can be represented
in terms of a linear combination of the matrix elements of the
projection operator (\ref{tf7})
\begin{equation}
\bigl(\Lambda_1\gamma_1\,\Lambda_2\gamma_2|s\Lambda_3\gamma_3\bigr)=
\sum_{\gamma_2'}^{}C(\gamma_2')\,\bigl<\Lambda_1\gamma_1\bigr|
\bigl<\Lambda_2\gamma_2\bigr|P_{\!\gamma_3,h}^{\Lambda_3}
\bigr|\Lambda_1h\bigl>\bigr|\Lambda_2\gamma_2'\bigr>~.
\label{CGC2}
\end{equation}
Classification of the multiple representations $\Lambda_3$ in the
representation $\Lambda_1\otimes\Lambda_2$ is a special problem
and we shall not touch it here.

We give here an explicit expression for the more general matrix
elements in comparison with the right-side of (\ref{CGC2}):
\begin{equation}
\bigl<\Lambda_1\gamma_1\bigr|\bigl<\Lambda_2\gamma_2\bigr|
P_{\!\gamma_3^{},\gamma_3'}^{\Lambda_3}
\bigr|\Lambda_1\gamma_1'\bigr>\bigr|\Lambda_2\gamma_2'\bigr>~.
\label{CGC3}
\end{equation}
Using (\ref{tf7})--(\ref{tf9}) and the Wigner-Racah calculus for
the subalgebra $\mathfrak{su}(2)$ it is not hard to obtain the
following result (see \cite{PST1,AST5}):
\begin{equation}
\begin{array}{l}
\bigl<\Lambda_1\gamma_1\bigr| \bigl<\Lambda_2\gamma_2\bigr|
P_{\gamma_3,\gamma_3'}^{\Lambda_3}
\bigr|\Lambda_1\gamma_1'\bigr>\bigr|\Lambda_2\gamma_2'\bigr>
=\bigl(t_1t_{1z}\,t_2t_{2z}\bigr|t_3t_{3z}\bigr)\;
\bigl(t_1t_{1z}'\,t_2t_{2z}'\bigr|t_3't_{3z}'\bigr) 
\\[14pt]
\quad\times(\lambda_3+1)(\mu_3+1)(\lambda_3+\mu_3+2)\;\;
A\!\sum\limits_{j_1'\!'j_2'\!'t_1'\!'t_2'\!'t_3'\!'}
C_{j_1'\!'j_2'\!'t_1'\!'t_2'\!'t_3'\!'}
\\[17pt]
\quad\times \left\{\!\!\!\!\!\!\!\!\!\begin{array}{cccc}
&j_1\!\!-\!j_1'\!'&\!\!j_2\!\!-\!j_2'\!'&\!\!
j_1\!\!+\!j_2\!\!-\!j_1'\!'\!\!-\!j_2'\!'\\
&t_1'\!'&t_2'\!'&t_3'\!'\\
&t_1&t_2&t_3\\
\end{array}
\!\!\!\right\}\!
\left\{\!\!\!\!\!\!\!\!\!\begin{array}{cccc}
&j_1'\!\!-\!j_1'\!'&j_2'\!\!-\!j_2'\!'&
j_1'\!\!+\!j_2'\!\!-\!j_1'\!'\!\!-\!j_2'\!'\\
&t_1'\!'&t_2'\!'&t_3'\!'\\
&t_1'&t_2'&t_3'
\end{array}
\!\!\!\right\},
\end{array}
\label{CGC4}
\end{equation}
where
\begin{equation}
\begin{array}{r}
A=\mbox{\large$\left(\frac{(2t_1+1)(2t_2+1)(2j_1+1)!(2j_2+1)!
(\lambda_3+\frac{1}{2}\mu_3-j_3+t_3+1)!
(\lambda_3+\frac{1}{2}\mu_3-j_3-t_3)!}
{(\lambda_1\!+\!\frac{1}{2}\mu_1\!-\!j_1\!+\!t_1\!+\!1)!
(\lambda_1\!+\!\frac{1}{2}\mu_1\!-\!j_1\!-\!t_1)!
(\lambda_2\!+\!\frac{1}{2}\mu_2\!-\!j_2\!+\!t_2\!+\!1)!
(\lambda_2\!+\!\frac{1}{2}\mu_2\!-\!j_2\!-\!t_2)!(2j_3)!}
\right.$}
\\[12pt]
\times\mbox{\large$\left.\frac{(2t_1'+1)(2t_2'+1)
(2j_1'+1)!(2j_2'+1)! (\lambda_3+\frac{1}{2}\mu_3-j_3'+t_3'+1)!
(\lambda_3+\frac{1}{2}\mu_3-j_3'-t_3')!}
{(\lambda_1\!+\!\frac{1}{2}\mu_1\!-\!j_1'\!+\!t_1'\!+\!1)!
(\lambda_1\!+\!\frac{1}{2}\mu_1\!-\!j_1'\!-\!t_1')!
(\lambda_2\!+\!\frac{1}{2}\mu_2\!-\!j_2'\!+\!t_2'+1)!
(\lambda_2\!+\!\frac{1}{2}\mu_2\!-\!j_2'\!-\!t_2')!(2j'_3)!}
\right)^{\frac{1}{2}}$},
\end{array}
\label{CGC5}
\end{equation}
\begin{equation}
\begin{array}{l}
C_{j_1'\!'j_2'\!'t_1'\!'t_2'\!'t_3'\!'}=\mbox{\large$\frac{
(-1)^{2(j_1\!+j_2\!+j_3'\!-j_1'\!'\!-j_2'\!')}
(2(j_1+j_2-j''_1-j''_2)+1)! 
}{(2j''_1)!(2j''_2)!(2j_1-2j''_1)!(2j_2-2j''_2)!
(2j_1'-2j''_2)!
}$}
\\[14pt]
\quad\times\mbox{\large$\frac{
(2(j_1'+j_2'-j''_1-j''_2)+1)!(2t''_1\!+\!1)(2t''_2\!+\!1)(2t''_3\!+\!1)}
{(2j_2'-2j''_2)!(2(j_1\!+\!j_2-\!j_3-\!j''_1\!-\!j''_2))!}$}
\\[14pt]
\quad\times\mbox{\large$\frac{
(\lambda_1\!+\frac{1}{2}\mu_1\!-\!j''_1\!+t''_1\!+\!1)!
(\lambda_1\!+\frac{1}{2}\mu_1\!-\!j''_1\!-t''_1)!
(\lambda_2\!+\frac{1}{2}\mu_2\!-\!j''_2\!+t''_2\!+\!1)!
(\lambda_2\!+\frac{1}{2}\mu_2\!-\!j''_2\!-t''_2)!
}{(\lambda_3+\frac{1}{2}\mu_3+j_1\!+\!j_2-\!j_3\!-\!
j''_1\!-\!j''_2\!+t''_3\!+\!2)!
(\lambda_3+\frac{1}{2}\mu_3+j_1\!+\!j_2-\!j_3\!-\!
j''_1\!-\!j''_2\!-\!t''_3\!+\!1)!}$}
\\[14pt]
\quad\times
{\displaystyle\left\{{j_1\!\!-\!\!j''_1\atop\frac{1}{2}\mu_1}\;
{j''_1\atop t_1}\;{j_1\atop t''_1}\right\}}
{\displaystyle\left\{{j_2\!\!-\!\!j''_2\atop\frac{1}{2}\mu_2}\;
{j''_2\atop t_2}\;{j_2\atop t''_2}\right\}}
{\displaystyle\left\{{j_3\atop t''_3}
{\,{j_1\!\!+\!\!j_2\!\!-\!\!j_3\!\!-\!\!j''_1\!\!-\!\!j''_2\atop
t_3}}\;
{\,{j_1\!\!+\!\!j_2\!\!-\!\!j''_1\!\!-\!\!j''_2\atop\frac{1}{2}\mu_3}}
\right\}}
\\[14pt]
\quad\times
{\displaystyle\left\{{j_1'\!\!-\!\!j''_1\atop\frac{1}{2}\mu_1}\;
{j''_1\atop t_1'}\;{j_1'\atop t''_1}\right\}}
{\displaystyle\left\{{j_2'\!\!-\!\!j''_2\atop\frac{1}{2}\mu_2}\;
{j''_2\atop t_2'}\;{j_2'\atop t''_2}\right\}}
{\displaystyle\left\{{j_3'\atop t''_3}\;{\,{j_1'\!\!+
\!j_2'\!\!-\!\!j_3'\!\!-\!\!j''_1\!\!-\!\!j''_2\atop t_3'}}
{\,{j_1'\!\!+\!\!j_2'\!\!-\!\!j''_1\!\!-\!\!j''_2
\atop\frac{1}{2}\mu_3}}\right\}}
\end{array}
\label{CGC6}
\end{equation}
Here everywhere the braces denote $6j$- and $9j$-symbols of the Lie
algebra $\mathfrak{su}(2)$.

 \setcounter{equation}{0}
\section{Bibliographical notes of applications for the extremal
projectors}

For the convenience of the reader the main development of the
subject will be characterized in this Section by the most
important references. In particular:
\begin{itemize}
\item
Explicit description of irreducible representations of
(super)algebras (construction of different bases, actions of
generators and their properties). \\[2pt] {\it Results}:
The Gel'fand-Tsetlin bases for:\\ $\mathfrak{su}(n)$ (R.M.~Asherova,
Yu.F.~Smir\-nov and V.N.~Tolstoy (1973)),\\ $\mathfrak{so}(n)$
(V.N.~Tolstoy (1975, unpublished)),\\ $G_2$ (D.T.~Svi\-ridov,
Yu.F.~Smir\-nov and V.N.~Tolstoy (1976)),\\ $\mathfrak{osp}(1|2)$
(F.A.~Berezin and  V.N.~Tolstoy (1980)),\\ $\mathfrak{gl}(m|n)$
(V.N.~Tolstoy, I.F.~Istomina and Yu.F.~Smirnov (1986)),\\
$U_q(\mathfrak{su}(n))$ (V.N.~Tols\-toy (1990)),\\
$U_q(\mathfrak{su}(1|n))$ (T.D.~Palev and V.N.~Tolstoy (1991)),\\
$\mathfrak{sp}(2n))$ (A.I.~Molev (1999)).
\medskip

\item
The theory of Clebsch-Gordan coefficients of the simple Lie
algebras. \\[2pt] {\it Results}:\\ $\mathfrak{su}(3)$ (Z.~Pluhar,
Yu.F.~Smirnov and V.N.~Tolstoy (1981-86)),\\ $U_q(\mathfrak{su}(2))$
(Yu.F.\,Smirnov, V.N.\,Tolstoy and Yu.I.\,Khatritonov (1991-93)),\\
$U_q(\mathfrak{su}(3))$ (R.M.~Ashe\-rova, Yu.F.~Smirnov
V.N.~Tolstoy (2001)),\\ $U_q(\mathfrak{su}(n))$ (V.N.~Tolstoy and
D.J.~Draayer (2000)).
\medskip

\item
Description of reduction algebras (Mikelson's algebras). \\[2pt]
{\it Results}: $A_n$, $B_n$, $C_n$, $D_n$ (D.P.~Zhelobenko (1983)),\\
$\mathfrak{su}(m|n)$, $\mathfrak{osp}(m|2n)$ (V.N.~Tolstoy
(1986)),\\ $U_q(\mathfrak{su}(n))$ (V.N. Tolstoy (1990)),\\
$U_q(\mathfrak{su}(1|n))$ (T.D.\ Palev and V.N.~Tolstoy (1991)).
\medskip

\item
Description of Verma modules of Lie (super)algebras (singular
vectors and their properties).\\[2pt] {\it Results}: for the simple Lie
algebras (D.P.~Zhelobenko (1985)).
\medskip

\item
Construction of solutions of the Yang-Baxter equation with the help of
projection operators. \\[2pt] {\it Results}: for $\mathfrak{u}(3)$ and
$\mathfrak{u}(n)$ (Yu.F.~Smirnov and V.N.~Tolstoy (1990);
V.\ Ta\-ra\-sov and A.~Varchenko (2002)).
\medskip

\item
Connection between extremal projectors and integral projection operators.\\[2pt]
{\it Results}: for the simple Lie algebras (A.N.~Leznov and
M.V.~Savel'ev (1974)).
\medskip

\item
Connection between extremal projectors and canonical elements.\\[2pt] {\it
Results}: for $q$-boson Kashiwara algebras (T.~Nakashima (2004)).
\medskip

\item
Generalization of extremal projectors.\\[2pt] {\it Results}: for
$\mathfrak{sl}(2)$ (V.N.~Tolstoy (1988)),\\ $U_q(\mathfrak{sl}(2))$
(H.-D.~Doebner and V.N.~Tolstoy (1996)).
\medskip

\item
Construction of indecomposable representations.\\[2pt] {\it Results}: for
$U_q(\mathfrak{sl}(2))$ (H.-D.~Doebner and V.N.~Tolstoy (1996)).
\end{itemize}
\medskip

 \def\acam  {Acta\,Appl.\,Math.}
 \def\comp  {Com\-mun.\,Math.\,Phys.}
 \def\czjp  {Czech.\,J.\,Phys.}
 \def\dakns {Dokl.\,Akad.\,Nauk.\,SSSR}
 \def\daknS {Dokl.\,Akad.\linebreak[0]Nauk.\,SSSR}
 \def\fuaa  {Funct.\linebreak[0]Anal.\,Appl.}
 \def\fuap  {Funkts.\,Anal.\,Pri\-lozh.}
 \def\izak  {Iz\-ves\-ti\-ya\,Akad.\,Nauk.\,SSSR\,Ser.\,Mat.}
 \def\jomp  {J.\,Math.\linebreak[0]Phys.}
 \def\jopa  {J.\,Phys.\linebreak[0]A:\,Math.\,Gen.}
 \def\maui  {Math.\,USSR\,Izv.}
 \def\nupb  {Nucl.\,Phys.~B}
 \def\remp  {Rev.\,Mod.\,Phys.}
 \def\romp  {Rep.\,Math.\,Phys.}
 \def\rums  {Russ.\,Math.\,Sur\-veys}
 \def\sjnp  {Sov.\,J.\,Nucl.\,Phys.}
 \def\slnp  {Sprin\-ger\,Lec\-ture\,No\-tes\,in\,Phy\-sics}
 \def\temf  {Te\-or.\,Mat.\,Fiz.}
 \def\thmp  {The\-or.\,Math.\,Phys.}
 \def\usmn  {Us\-pekhi\,Mat.\,Nauk}
 \def\yafi  {Ya\-der\-na\-ya\,Fiz.}

\medskip
\end{document}